\documentclass[conference]{IEEEtran}
\usepackage{cite}

\ifCLASSINFOpdf
\else
\usepackage[dvips]{graphicx}
\fi
\usepackage[cmex10]{amsmath}
\begin{document}
%
\title{On Max-SINR Receiver for Hexagonal Multicarrier Transmission Over Doubly Dispersive Channel}

\author{\IEEEauthorblockN{Kui~Xu,~Youyun~Xu,~Xiaochen~Xia,~Dongmei Zhang}
\IEEEauthorblockA{Institute of Communications Engineering\\PLA University of Science and Technology\\Nanjing 210007, P. R. China.\\
Email:lgdxxukui@126.com, yyxu@vip.sina.com}}


%


\maketitle

\begin{abstract}
In this paper, a novel receiver for Hexagonal Multicarrier
Transmission (HMT) system based on the maximizing
Signal-to-Interference-plus-Noise Ratio (Max-SINR) criterion is
proposed. Theoretical analysis shows that the prototype pulse of the
proposed Max-SINR receiver should adapt to the root mean square
(RMS) delay spread of the doubly dispersive (DD) channel with
exponential power delay profile and U-shape Doppler spectrum.
Simulation results show that the proposed Max-SINR receiver
outperforms traditional projection scheme and obtains an
approximation to the theoretical upper bound SINR performance within
the full range of channel spread factor. Meanwhile, the SINR
performance of the proposed prototype pulse is robust to the
estimation error between the estimated value and the real value of
time delay spread.
\end{abstract}


%
\IEEEpeerreviewmaketitle

\section{Introduction}
Orthogonal frequency division multiplexing (OFDM) systems with
guard-time interval or cyclic prefix can prevent inter-symbol
interference (ISI). OFDM has overlapping spectra and rectangular
impulse responses. Consequently, each OFDM sub-channel exhibits a
sinc-shape frequency response. Therefore, the time variations of the
channel during one OFDM symbol duration destroys the orthogonality
of different subcarriers, and results in power leakage among
subcarriers, known as inter-carrier interference (ICI), which causes
degradation in system performance. In order to overcome the above
drawbacks of OFDM system, several pulse-shaping OFDM systems were
proposed \cite{Kumb07,Das07,Abb10,Gao11}.

It is shown that signal transmission through a rectangular lattice
is suboptimal for doubly dispersive (DD) channel
\cite{Str03,Han07,Han09}. By using results from sphere covering
theory, the authors have demonstrated that lattice OFDM (LOFDM)
system, which is OFDM system based on hexagonal-type lattice,
providing better performance against ISI/ICI \cite{Str03}. However,
LOFDM confines the transmission pulses to an orthogonal set. As
pointed out in \cite{Das07}, these orthogonalized pulses destroy the
time-frequency (T-F) concentration of the initial pulses, hence
lower the robustness to the time and frequency dispersion caused by
the DD propagation channel.

In \cite{Han07,Han09,Han10}, the authors abandoned the orthogonality
condition of the modulated pulses and proposed a multicarrier
transmission scheme on hexagonal T-F lattice named as hexagonal
multicarrier transmission (HMT). To optimally combat the impact of
the DD propagation channels, the lattice parameters and the pulse
shape of modulation waveform are jointly optimized to adapt to the
channel scattering function from a minimum energy perturbation point
of view. It is shown that the hexagonal multicarrier transmission
systems obtain lower energy perturbation by incorporation the best
T-F localized Gaussian pulses as the elementary modulation waveform,
hence outperform OFDM and LOFDM systems from the robustness against
channel dispersion point of view \cite{Xu09,Xu11,Xu12}.

In HMT system, there is no cyclic prefix and the data symbols of HMT
signal are transmitted at the hexagonal lattice point of T-F plane
and subcarriers are interleaved in the T-F space, as is illustrated
in Fig. 1. The basic mathematical operation on the received signal
performed by the demodulator is a projection onto an identically
structured function set generated by the prototype pulse function
\cite{Xu11,Xu12,Jun07}, i.e. an optimal match filter. It is shown in
\cite{Wu05,Wu07,Wu071} that the optimum sampling time of wireless
communication systems over DD channel depends on the power
distribution of the channel profiles, and that zero timing offset
does not always yield the best system performance. Max-SINR
ISI/ICI-Shaping receiver for multicarrier modulation system is
discussed in \cite{Das07}, the Max-SINR prototype pulse can be
obtained by maximizing the generalized Rayleigh quotient.

We will present in this paper that receivers proposed in
\cite{Han07,Han09,Han10} are suboptimal approaches in the view of
Signal-to-Interference-plus-Noise Ratio (SINR) performance. A novel
receiver based on maximizing SINR (Max-SINR) criterion for HMT
system is proposed in this paper. Theoretical analyses and
simulation results show that the proposed optimal Max-SINR receiver
outperforms the traditional projection receiver in
\cite{Han07,Han09,Han10} and obtains an approximation to the
theoretical upper bound SINR performance. Meanwhile, the proposed
scheme is robust to the estimation error between the estimated value
and the real value of root mean square (RMS).

\section{Hexagonal Multicarrier Transmission System}
In the view of signal transmission on lattice in the T-F plane, the
system performance is mainly determined by two factors: a) energy
concentration of the elementary modulation pulse, a better T-F
concentrated pulse would lead to more robustness against the energy
leakage and b) distance between the transmitted symbols in the T-F
plane, it is obvious that the larger the distance, the less the
perturbation among the transmitted symbols. As pointed out in
\cite{Han07,Han09,Han10}, for a given signaling efficiency, the
information-bearing pulses arranged on a hexagonal T-F lattice, as
is illustrated in Fig. 1, can be separated as sufficiently as
possible in the T-F plane.
\begin{figure}[!t] \centering
\centering
\includegraphics[width=3.2in]{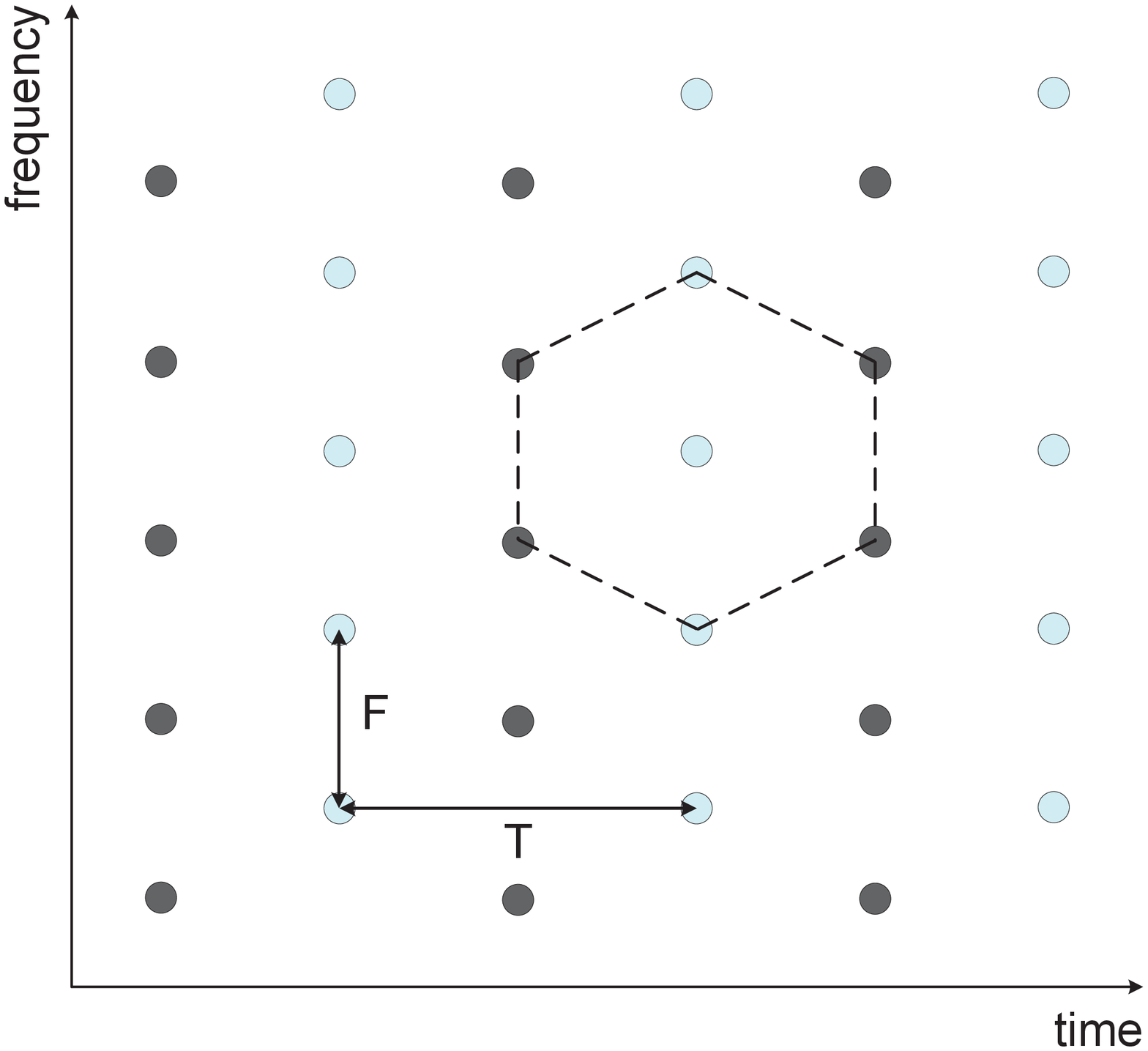}
\caption{Partition of the hexagonal lattice into a rectangular
sublattice $\textit{\textbf{V}}_{\textrm{rect}1}$ (denote by
$\circ$) and its coset $\textit{\textbf{V}}_{\textrm{rect}2}$
(denote by $\bullet$).} \label{fig_sim}
\end{figure}

In HMT systems, the transmitted baseband signal can be expressed as
\cite{Han07}

\begin{equation} \label{1}
\begin{split}
x(t)&=\sum_{m}\sum_{n}c_{m,2n}g\big(t-mT\big)e^{j2\pi nFt}\\
&+\sum_{m}\sum_{n}c_{m,2n+1}g\big(t-mT-\frac{T}{2}\big)e^{j2\pi(nF+\frac{F}{2})t}
\end{split}
\end{equation}
where $T$ and $F$ are the lattice parameters, which can be viewed as
the symbol period and the subcarrier separation, respectively;
$c_{m,n}$ is the user data, which is assumed to be taken from a
specific signal constellation and independent and identically
distributed (i.i.d.) with zero mean and average power
$\sigma_{c}^{2}$; $m\in\mathcal {M}$ and $n\in\mathcal {N}$ are the
position index in the T-F plane; $\mathcal {M}$ and $\mathcal {N}$
denote the sets from which $m,n$ can be taken, with cardinalities
$\textit{M}$ and $\textit{N}$, respectively. The prototype pulse
$g(t)$ is the Gaussian window

\begin{equation} \label{2}
g(t)=(2/\sigma)^{1/4}e^{-(\pi/\sigma)t^{2}}
\end{equation}
with $\sigma$ being a parameter controlling the energy distribution
in the time and frequency directions. The ambiguity function of
Gaussian pulse is defined by

\begin{equation} \label{3}
\begin{split}
A_{g}(\tau,\upsilon)&=\int_{-\infty}^{\infty}g(t)g^{*}(t-\tau)e^{-j2\pi\upsilon
t}dt \\
 &=e^{-\frac{\pi}{2}(\frac{1}{\sigma}\tau^{2}+\upsilon^{2})}e^{-j\pi\tau\upsilon}
\end{split}
\end{equation}
where $(\cdot)^{*}$ denotes the complex conjugate.

It is shown in Fig. 1 that the original hexagonal lattice can be
expressed as the disjoint union of a rectangular sublattice
$\textit{\textbf{V}}_{\textrm{rect}1}$ and its coset
$\textit{\textbf{V}}_{\textrm{rect}2}$. The transmitted baseband
signal in (1) can be rewritten as
\begin{equation} \label{4}
x(t)=\sum_{i}\sum_{m}\sum_{n}c_{m,n}^{i}g_{m,n}^{i}(t)
\end{equation}
where $i$=$1,2$, $c_{m,n}^{1}$ and $c_{m,n}^{2}$ represent the
symbols coming from two sets $\textit{\textbf{V}}_{\textrm{rect}1}$
and $\textit{\textbf{V}}_{\textrm{rect}2}$, respectively.
$g_{m,n}^{i}(t)=g(t-mT-\frac{iT}{2})e^{j2\pi(nF+\frac{iF}{2})t}$ is
the transmitted pulse on $m$-th symbol and $n$-th subcarrier.

The baseband DD channel can be modeled as a random linear operator
$\textrm{H}$ \cite{Bel63}
\begin{equation} \label{5}
\textrm{H}[x(t)]=\int_{0}^{\tau_{\textrm{max}}}\int^{f_{d}}_{-f_{d}}H(\tau,\upsilon)x(t-\tau)e^{j2\pi\upsilon
t}d\tau d\upsilon
\end{equation}
where $\tau_{\textrm{max}}$ and $f_{d}$ are the maximum multipath
delay spread and the maximum Doppler frequency,
respectively\cite{Coh95}. $H(\tau,\upsilon)$ is called the
delay-Doppler spread function, which is the Fourier transform of the
time-varying channel impulse response $h(t,\tau)$ with respect to
$t$. The product $\vartheta=\tau_{\textrm{max}}f_{d}$ is referred to
as the channel spread factor (CSF). If $\vartheta<1$, the channel is
said to be underspread; otherwise, it is overspread. Practical
wireless channels usually satisfy the assumption of wide-sense
stationary uncorrelated scattering (WSSUS), and $\vartheta\ll1$
\cite{Bel63}.

In the WSSUS assumption the channel is characterized by the second
order statistics
\begin{equation} \label{6}
E[H(\tau,\upsilon)H^{*}(\tau_{1},\upsilon_{1})]=S_{H}(\tau,\upsilon)\delta(\tau-\tau_{1})\delta(\upsilon-\upsilon_{1})
\end{equation}
where $E[\cdot]$ denotes the expectation and $S_{H}(\tau,\upsilon)$
is called the scattering function, which characterizes the
statistics of the WSSUS channel. Without loss of generality, we use
$\int_{0}^{\tau_{\textrm{max}}}\int_{-f_{d}}^{f_{d}}S_{H}(\tau,\upsilon)d\tau
d\upsilon=1$.

As shown in (5), the propagation channel introduces energy
perturbation among the transmitted symbols. It is shown in
\cite{Str03,Han07,Han09} that the symbol energy perturbation
function is dependent on the channel scattering function and the
pulse shape. The properly designed HMT system with Gaussian
prototype pulse achieves minimum symbol energy perturbation over DD
channel. The optimality of such system in combating the ISI/ICI
caused by the DD channel is guaranteed by the Heisenberg uncertainty
principle and the sphere-packing theory. The choice of the system
parameters $(\sigma, T, F)$ to minimize such undesired energy
perturbation is extensively dealt with in \cite{Str03,Han07,Han09}.
In general, the pulse shape of the Gaussian window $\sigma$ and the
transmission pattern parameters $T$ and $F$ should be matched to the
channel scattering function. The optimal system parameter for DD
channels with exponential-U scattering function can be chosen as
\cite{Han07}\footnotemark \footnotetext{In the case of no time and
frequency dispersion,
$S_H(\tau,\upsilon)=\delta(\tau)\delta(\upsilon)$, the matching
criteriaes (7) and (8) are also reduced to
$(\sqrt{3}T)/(F)=(W_t)/(W_f)$ and $T/(\sqrt{3}F)=(W_t)/(W_f)$,
respectively. $W_t^2$ and $W_f^2$ are the centralized temporal and
spectral second-order moments, respectively \cite{Coh95}.}
\begin{equation} \label{7}
\sigma=\alpha\frac{\tau_{\textrm{rms}}}{f_{d}}=\sqrt{3}\frac{T}{F}
\end{equation}
and
\begin{equation} \label{8}
\sigma=\alpha\frac{\tau_{\textrm{rms}}}{f_{d}}=\frac{1}{\sqrt{3}}\frac{T}{F}
\end{equation}
where $\tau_{\textrm{rms}}$ is the RMS delay of the DD channel  and
the coefficient $\alpha$ for various signaling efficiency $\rho$ is
listed in Table I.
\begin{table}[!t]
\renewcommand{\arraystretch}{1.3}
\caption{Factor $\alpha$ for Various Signaling Efficiencies}
\label{table_example} \centering
\begin{tabular}{|c|c|c|c|c|}
\hline
\bfseries $\rho$ &  0.5 &  1.0 &  2.0 &  4.0\\
\hline
$\alpha$ & 2.25 & 2.00 & 1.90 & 1.85\\
\hline
\end{tabular}
\end{table}
The received signal can be expressed as
\begin{equation} \label{9}
r(t)=\textrm{H}[x(t)]+w(t)
\end{equation}
where $w(t)$ is the AWGN with variance $\sigma_{w}^{2}$.

\section{Max-SINR Receiver for HMT System}
The basic mathematical operation of the received signal performed by
the demodulator is a projection onto an identically structured
function set generated by the prototype pulse function, i.e. an
optimal match filter\cite{Han09,Xu12}. To obtain the data symbol
$\hat{c}_{m,n}^{i}$, the match filter receiver projects the received
signal $r(t)$ on the prototype pulse function set
$\psi_{m,n}^{i}(t),i$=$1,2$, i.e.,
\begin{equation} \label{10}
\begin{split}
\hat{c}_{m,n}^{i}&=\big<r(t),\psi_{m,n}^{i}(t)\big> \\
 &=\sum_{j}\sum_{m',n'}c_{m',n'}^{j}\big<\textrm{H}[g_{m',n'}^{j}(t)],\psi_{m,n}^{i}(t)\big>\\
 &+\big<w(t),\psi_{m,n}^{i}(t)\big>
\end{split}
\end{equation}
where $\psi_{m,n}^{i}(t)$=$\psi(t-mT-\frac{i}{2}T)e^{j2\pi
(nF+\frac{iF}{2})t}$, and $\psi(t)$ is the prototype pulse at the
receiver. The energy of the received symbol, after projection on the
prototype pulse set $\psi_{m,n}^{i}(t)$ can be expressed as
\begin{equation} \label{11}
\begin{split}
E_{s}&=\textrm{E}\Big\{\Big|\sum_{m',n'}c_{m',n'}^{j}\big<\textrm{H}[g_{m',n'}^{j}(t)],\psi_{m,n}^{i}(t)\big>\\
&+\big<w(t),\psi_{m,n}^{i}(t)\big>\Big|^2\Big\}
\end{split}
\end{equation}
Under the assumptions of WSSUS channel and source symbols are
statistically independent, (11) can be rewritten as
\begin{equation} \label{12}
\begin{split}
E_{s}&=\sigma_{c}^{2}\int_{\tau}\int_{\upsilon}S_{H}(\tau,\upsilon)\\
&\cdot\bigg[\sum_{m,n}\bigg(\big|A_{g,\psi}(mT+\tau, nF+\upsilon)\big|^{2} \\
&+\big|A_{g,\psi}\big(mT+\frac{T}{2}+\tau,nF+\frac{F}{2}+\upsilon\big)\big|^2\bigg)\bigg]d\tau
 d\upsilon\\
&+\sigma_{w}^{2}\big|A_{g,\psi}(0,0)\big|
\end{split}
\end{equation}

The SINR of received signal can be expressed as
\begin{equation} \label{13}
R_{\textrm{SIN}}=\frac{\sigma_{c}^{2}}{E_{\textrm{IN}}}\int_{\tau}\int_{\upsilon}S_{H}(\tau,\upsilon)\big|A_{g,\psi}(\tau,\upsilon)\big|^2d\tau
d\upsilon
\end{equation}
where the interference-plus-noise energy
\begin{equation} \label{14}
\begin{split}
E_{\textrm{IN}}&=\sigma_{c}^{2}\int_{\tau}\int_{\upsilon}S_{H}(\tau,\upsilon)\\
&\cdot\bigg[\sum_{z=[m,n]^{T}\neq[0,0]^{T}}\bigg(\big|A_{g,\psi}(mT+\tau, nF+\upsilon)\big|^{2} \\
 &+\big|A_{g,\psi}\big(mT+\frac{T}{2}+\tau,nF+\frac{F}{2}+\upsilon\big)\big|^2\bigg)\bigg]d\tau
 d\upsilon\\
&+\sigma_{w}^{2}\big|A_{g,\psi}(0,0)\big|
\end{split}
\end{equation}

Clearly, the interference-plus-noise energy function
$R_{\textrm{IN}}$ is dependent on the channel scattering function
and the pulse shape (through its ambiguity function). According to
the form of channel scattering functions, the Max-SINR receiver can
be discussed in two cases \cite{Han07}: Case A: DD channel with
exponential power delay profile and U-shape Doppler spectrum. Case
B: DD channel with uniform power delay profile and uniform Doppler
spectrum. Due to the space limitations, this paper only to discuss
the Case A, that is the Max-SINR receiver for HMT system over DD
channel with exponential power delay profile and U-shape Doppler
spectrum.

For the DD channel with exponential power delay profile and U-shape
Doppler spectrum, the scattering function can be expressed as
\cite{Mat02}
\begin{equation} \label{15}
S_{H}(\tau,\upsilon)=\frac{e^{\frac{-\tau}{\tau_{\textrm{rms}}}}}{\pi\tau_{\textrm{rms}}f_{d}\sqrt{1-(\upsilon/f_{d})^2}}
\end{equation}
with $\tau>0,|\upsilon|<f_{d}$. We assume that \footnotemark
\footnotetext{It is shown in \cite{Wu05,Wu07,Wu071} that the optimum
sampling time of wireless communication systems over DD channel
depends on the power distribution of the channel profiles, and that
zero timing offset does not always yield the best system
performance. In other words, there is a timing offset between the
prototype pulses at the transmitter and the receiver. In
\cite{Das07}, the Max-SINR prototype pulse $\textit{\textbf{g}}$ of
multicarrier transmission system with rectangular T-F lattice over
DD channel is obtained by maximizing the generalized Rayleigh
quotient
$\hat{\textit{\textbf{g}}}=\arg\max\frac{\textit{\textbf{g}}^H\textit{\textbf{B}}\textit{\textbf{g}}}{\textit{\textbf{g}}^H\textit{\textbf{A}}\textit{\textbf{g}}}$.
The solution is the generalized eigenvector of the matrix pair
$(\textit{\textbf{B}},\textit{\textbf{A}})$ corresponding to the
largest generalized eigenvalue. It is shown that there is a delay
between the transmitted Gaussian prototype pulse and the received
Gaussian prototype pulse. In this paper, the close form time offset
expressions between the transmitted and received prototype pulse is
derived.} $\psi(t)=g(t$-$\Delta t)$, the theoretical SINR of the
received signal over the DD channel with exponential power delay
profile and U-shape Doppler spectrum can be expressed as
\begin{equation} \label{16}
\begin{split}
R_{\textrm{SIN}}&=\frac{\sigma_{c}^{2}}{\pi\tau_{\textrm{rms}}f_{d}E_{\textrm{IN}}}\int_{0}^{\infty}e^{-\frac{\tau}{\tau_{\textrm{rms}}}}e^{-\frac{\pi}{\sigma}(\tau-\Delta t)^{2}}d\tau\\
&\cdot\int_{-f_{d}}^{f_{d}}\frac{e^{-\sigma\pi\upsilon^{2}}}{\sqrt{1-(\upsilon/f_{d})^{2}}}d\upsilon
\end{split}
\end{equation}
Substituting (15) into (14), the interference-plus-noise energy
function $E_{\textrm{IN}}$ in (16) can be expressed as
\begin{equation} \label{17}
\begin{split}
E_{\textrm{IN}}&=\frac{\sigma_{c}^{2}}{\pi\tau_{\textrm{rms}}f_{d}}\Bigg\{\sum_{(m,n)\neq(0,0)}\int_{0}^{\infty}e^{-\frac{\tau}{\tau_{\textrm{rms}}}}e^{-\frac{\pi(mT+\tau-\Delta t)^{2}}{\sigma}}d\tau\\
&\cdot\int_{-f_{d}}^{f_{d}}\frac{e^{-\sigma\pi(nF+\upsilon)^{2}}}{\sqrt{1-(\upsilon/f_{d})^{2}}}d\upsilon\\
&+\sum_{(m,n)\neq(0,0)}\int_{0}^{\infty}e^{-\frac{\tau}{\tau_{\textrm{rms}}}}e^{-\frac{\pi\big(mT+\frac{T}{2}+\tau-\Delta t\big)^{2}}{\sigma}}d\tau\\
&\cdot\int_{-f_{d}}^{f_{d}}\frac{e^{-\sigma\pi\big(nF+\frac{F}{2}+\upsilon\big)^{2}}}{\sqrt{1-(\upsilon/f_{d})^{2}}}d\upsilon\Bigg\}+\sigma_{w}^{2}\big|A_{g,\psi}(0,0)\big|
\end{split}
\end{equation}

The theoretical SINR upper bound of the received signal can be
expressed as
\begin{equation}\label{18}
R_{\textrm{UB}}=\arg\max_{\Delta t}R_{\textrm{SIN}}
\end{equation}
Plugging (16) and (17) in (18), the Max-SINR prototype pulse can be
expressed as $\psi(t)=g(t-\Delta t)$ and (see Appendix)
\begin{equation} \label{20}
\begin{split}
\Delta
t&=\frac{\sigma}{2\pi\tau_{\textrm{rms}}}\\
&-\sqrt{\frac{\sigma}{2\pi}}\Bigg(\frac{\frac{3.28\sqrt{\sigma}}{\tau_{\textrm{rms}}}-\sqrt{\frac{3.28^2\sigma}{\tau_{\textrm{rms}}^2}-3.52\big(\frac{\sigma}{\tau^2_{\textrm{rms}}-4}}\big)}{1.76}\Bigg)
\end{split}
\end{equation}

We can see from equation (19) that the prototype pulse of the
proposed Max-SINR receiver is a function of RMS. In HMT system, the
prototype pulse shape of the Gaussian window and the transmission
pattern parameters $T$ and $F$ should be matched to the channel
scattering function including RMS. In other word, the RMS delay
spread and the maximum Doppler frequency are known at the HMT
transceiver. Hence, it is reasonable to assume that the RMS delay
spread priori known in HMT system.

\section{Simulation and Discussion}
In this section, we test the proposed Max-SINR receiver via computer
simulations based on the discrete signal model. In the following
simulations, the number of subcarriers for HMT system is chosen to
$N$=$40$, and the length of prototype pulse $N_{g}$=$600$. The
center carrier frequency is $f_{c}$=$5$GHz and the sampling interval
is set to $T_{s}$=$10^{-6}$s. The system parameters of HMT system
are $F$=$25$kHz, $T$=$1\times10^{-4}s$ and $\sigma$ is set to
$\sigma$=$T/\sqrt{3}F$. Traditional projection receiver proposed in
\cite{Han07,Han09,Han10} is named as Traditional Projection Receiver
(TPR) in the following simulation results. WSSUS channel is choosen
as DD channel with exponential power delay profile and U-shape
Doppler spectrum.

The prototype pulses of Max-SINR receiver at $\vartheta$=$0.1$ and
$\vartheta$=$0.04$ are given in Fig. 2. The prototype pulse of TPR
scheme is also depicted for comparison. We can see from Fig. 2 that
there is a delay between the TPR prototype pulse and the Max-SINR
prototype pulse. The delay is obtained according to equation (19),
and increases with the increasing of CSF $\vartheta$.

The SINR performance of different receivers with the variety of
$\sigma_{c}^{2}/\sigma_{w}^{2}$ for HMT system over DD channel is
depicted in Fig. 3. The CSFs are set to $\vartheta$=0.07 and
$\vartheta$=0.2, respectively. We can see from Fig. 3 that the SINR
performance of the proposed Max-SINR receiver outperforms TPR scheme
about 1$\sim4\textrm{dB}$ at $\vartheta$=0.07 and
1.5$\sim3.5\textrm{dB}$ at $\vartheta$=$0.2$, respectively. The SINR
gap between the proposed Max-SINR receiver and the theoretical SINR
upper bound is smaller than 0.5dB and 0.1dB at
$\vartheta$=$0.07,0.2$, respectively.

The SINR performance of the proposed Max-SINR receiver with the
variety of $\sigma_{c}^2/\sigma_{w}^2$ is given in Fig. 4. We assume
that there is an estimation error between the estimated value of RMS
delay spread $\tau'_{\textrm{rms}}$ and the real value
$\tau_{\textrm{rms}}$. We can conclude from Fig. 4 that the proposed
Max-SINR receiver is robust to the estimation error of RMS delay
spread.

The SINR performance with the variety of channel spread factors
$\vartheta$ at $\sigma_{c}^2/\sigma_{w}^2$=20dB is depicted in Fig.
5. The SINR performance of the TPR scheme is depicted for
comparison. It can be seen that there is a degradation of SINR with
the increasing of channel spread factor. The proposed Max-SINR
receiver obtains an approximation to the theoretical upper bound
SINR performance within the full range of $\vartheta$. There is an
about 2.5dB maximum SINR gap between the Max-SINR receiver and the
TPR scheme at $\vartheta$=0.35, and the SINR gap decreases as the
CSF $\vartheta$ decreases.
\begin{figure}[!t]
\centering
\includegraphics[width=3.5in]{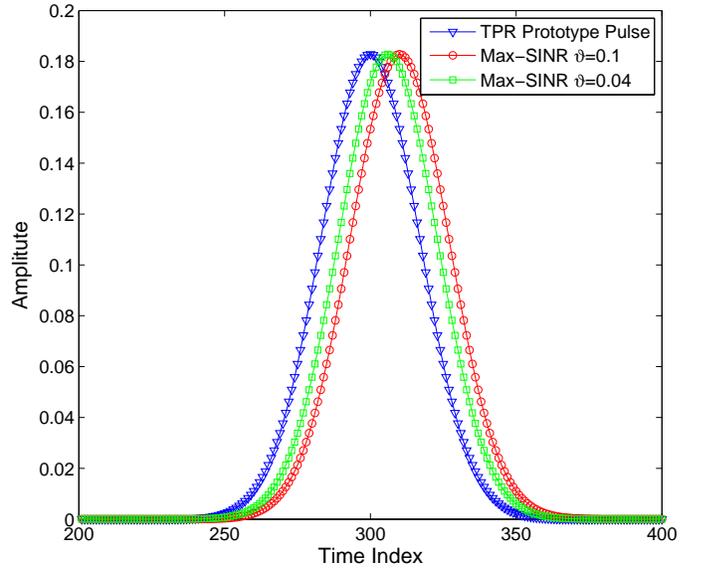}
\caption{The comparison of prototype pulses for different receivers
at $\vartheta$=0.1 and $\vartheta$=0.04, respectively.}
\label{fig_sim}
\end{figure}

\begin{figure}[!t]
\centering
\includegraphics[width=3.5in]{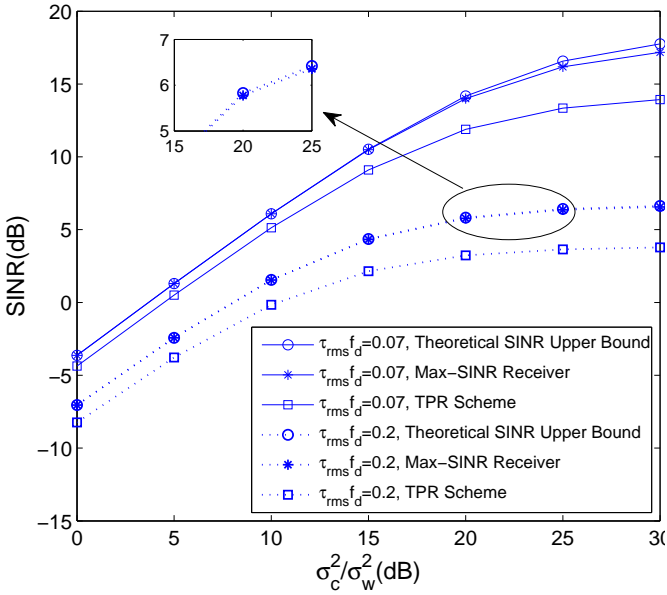}
\caption{The SINR performance of different receivers with the
variety of $\sigma_{c}^2/\sigma_{w}^2$ for HMT system over DD
channel.} \label{fig_sim}
\end{figure}

\begin{figure}[!t]
\centering
\includegraphics[width=3.5in]{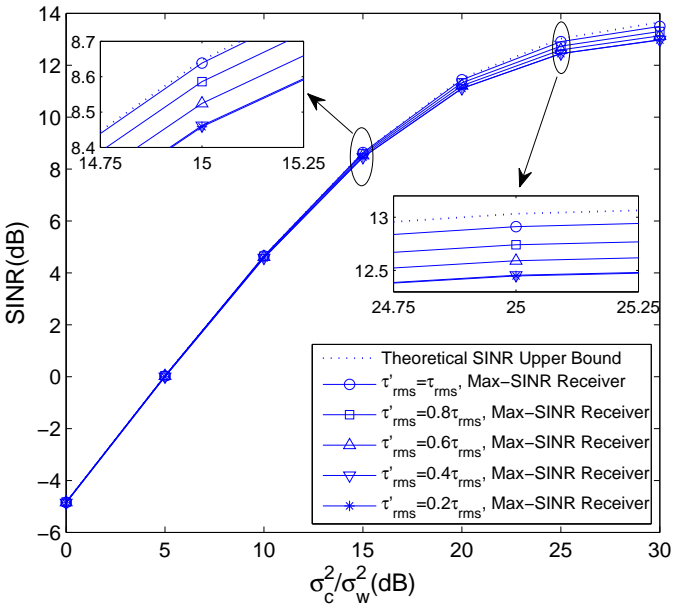}
\caption{The SINR performance of the proposed Max-SINR receiver for
HMT system over DD channel. We assume that there is an estimation
error between the estimated value $\tau'_{\textrm{rms}}$ and the
real value $\tau_{\textrm{rms}}$.} \label{fig_sim}
\end{figure}

\begin{figure}[!t]
\centering
\includegraphics[width=3.5in]{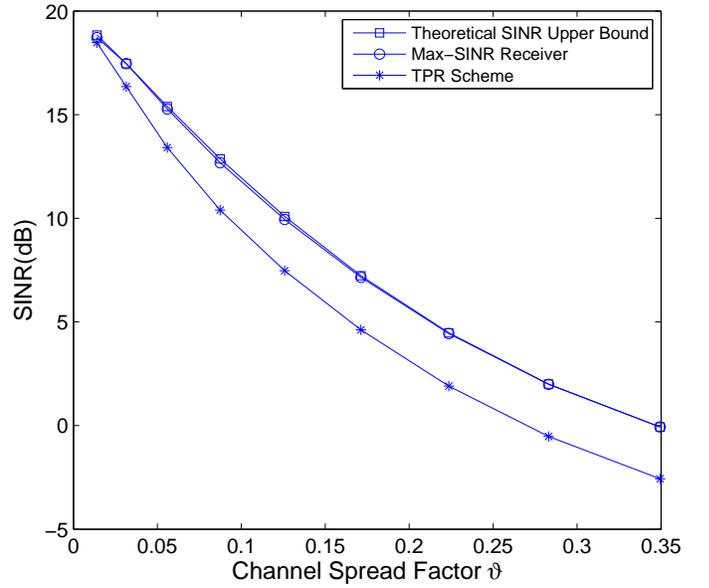}
\caption{The SINR performance of different receivers with the
variety of channel spread factors $\vartheta$,
$\sigma_{c}^2/\sigma_{w}^2$=20dB.} \label{fig_sim}
\end{figure}

\section{Conclusion}
The Max-SINR receiver is proposed for HMT system over DD channel in
this paper. After a detailed analysis, we present that the prototype
pulse of the Max-SINR receiver should adapt to the RMS delay spread.
Theoretical analysis shows that the proposed Max-SINR receiver
outperforms the TPR scheme in SINR and obtains an approximation to
the theoretical upper bound SINR performance within the full range
of CSF. Simulation results show that the proposed scheme is robust
to the estimation error of RMS and is suitable for the actual HMT
system.

\appendices

\section{Proof of Equation(19)}
Under the assumption that the prototype pulse at the receiver is
$\psi(t)=g(t-\Delta t)$, the SINR of the received signal can be
expressed as
\begin{equation} \label{3}
\begin{split}
R_{\textrm{SIN}}&=\frac{\sigma_{c}^{2}}{\pi\tau_{\textrm{rms}}f_{d}E_{\textrm{IN}}}\int_{0}^{\infty}e^{-\frac{\tau}{\tau_{\textrm{rms}}}}e^{-\frac{\pi}{\sigma}(\tau-\Delta
t)^{2}}d\tau\\
&\cdot\int_{-f_{d}}^{f_{d}}\frac{e^{-\sigma\pi\upsilon^{2}}}{\sqrt{1-(\upsilon/f_{d})^{2}}}d\upsilon
\\
&\simeq\frac{\sigma_{c}^{2}}{\sigma_{w}^{2}\tau_{\textrm{rms}}f_{d}}\int_{-f_{d}}^{f_{d}}\frac{e^{-\sigma\pi\upsilon^{2}}}{\sqrt{1-(\upsilon/f_{d})^2}}d\upsilon\\
&\cdot\Bigg(\underbrace{e^{\frac{\sigma}{4\pi\tau_{\textrm{rms}}^{2}}+\frac{\Delta
t}{\tau_{\textrm{rms}}}}}_{a(\Delta
t)}\underbrace{\int_{0}^{\infty}e^{-\frac{\pi}{\sigma}(\tau-\Delta
t+\frac{\sigma}{2\pi\tau_{\textrm{rms}}})^2}d\tau}_{b(\Delta
t)}\Bigg)
\end{split}
\end{equation}
where $b(\Delta t)$ in (20) can be rewritten as
\begin{equation} \label{3}
\begin{split}
b(\Delta
t)&=\sqrt{\frac{\sigma}{\pi}}\int^{\infty}_{\sqrt{\frac{\pi}{\sigma}}(\frac{\sigma}{2\pi\tau_{\textrm{rms}}}-\Delta
t)}e^{-x^2}dx\\
&=\frac{\sqrt{\sigma}}{2}\textrm{erfc}\Bigg(\sqrt{\frac{\pi}{\sigma}}\Big(\frac{\sigma}{2\pi\tau_{\textrm{rms}}}-\Delta
t\Big)\Bigg)
\end{split}
\end{equation}
where $\textrm{erfc}(\cdot)$ is the complementary error function. If
$x>0$, we may obtain an approximate solution of the complementary
error function $\textrm{erfc}(\cdot)$ by \cite{Kin05}
\begin{equation} \label{3}
\textrm{erfc}(\frac{x}{\sqrt{2}})\simeq\frac{2e^{-\frac{x^2}{2}}}{1.64x+\sqrt{0.76x^2+4}}
\end{equation}

The Max-SINR receiver can be obtained by solving the gradient of
$a(\Delta t)b(\Delta t)$ with respect to $\Delta t$
\begin{equation} \label{3}
\frac{da(\Delta t)}{d\Delta t}b(\Delta t)+\frac{db(\Delta
t)}{d\Delta t}a(\Delta t)=0
\end{equation}
where $\frac{da(\Delta t)}{d\Delta t}=-\frac{a(\Delta
t)}{\tau_{\textrm{rms}}}$ and $\frac{db(\Delta t)}{d\Delta t}$ can
be expressed as
\begin{equation} \label{3}
\frac{db(\Delta t)}{d\Delta
t}=e^{-\frac{\pi}{\sigma}\big(\frac{\sigma}{2\pi\tau_{\textrm{rms}}}-\Delta
t\big)^2}
\end{equation}
hence, equation (23) can be rewritten as
\begin{equation}
\begin{split}
\label{eqn_dbl_x} \frac{b(\Delta
t)}{\tau_{\textrm{rms}}}&=e^{-\frac{\pi}{\sigma}\big(\frac{\sigma}{2\pi\tau_{\textrm{rms}}}-\Delta
t\big)^2}\\
&=\frac{\sqrt{\sigma}}{2\tau_{\textrm{rms}}}\textrm{erfc}\bigg(\sqrt{\frac{\pi}{\sigma}}\big(\frac{\sigma}{2\pi\tau_{\textrm{rms}}}-\Delta
t\big)\bigg)\\
&\simeq\frac{\sqrt{\sigma}}{\tau_{\textrm{rms}}}e^{-\frac{\pi}{\sigma}\big(\frac{\sigma}{2\pi\tau_{\textrm{rms}}}-\Delta
t\big)^2}\Bigg(1.64\sqrt{\frac{2\pi}{\sigma}}\big(\frac{\sigma}{2\pi\tau_{\textrm{rms}}}\\
&-\Delta
t\big)+\sqrt{\frac{1.52\pi}{\sigma}\big(\frac{\sigma}{2\pi\tau_{\textrm{rms}}}-\Delta
t\big)^2+4}\Bigg)^{-1}
\end{split}
\end{equation}
Equation (25) can be simplified to a quadratic equation. Under the
constraint of $\Delta t>0$, the solution of the quadratic equation
can be expressed as
\begin{equation} \label{3}
\begin{split}
\Delta
t&=\frac{\sigma}{2\pi\tau_{\textrm{rms}}}\\
&-\sqrt{\frac{\sigma}{2\pi}}\Bigg(\frac{\frac{3.28\sqrt{\sigma}}{\tau_{\textrm{rms}}}-\sqrt{\frac{3.28^2\sigma}{\tau_{\textrm{rms}}^2}-3.52\big(\frac{\sigma}{\tau^2_{\textrm{rms}}-4}}\big)}{1.76}\Bigg)
\end{split}
\end{equation}

\section*{Acknowledgment}
This work was supported by the National Natural Science Foundation
of China (No. 60972050) and Major Special Project of China under
Grant (2010ZX03003-003-01) and the Jiangsu Province National Science
Foundation under Grant (BK2011002) and the Young Scientists
Pre-research Fund of PLAUST under Grant (No. KYTYZLXY1211).



%

\end{document}